\documentclass[aps,prd,12pt,nofootinbib]{revtex4}

\usepackage{amsfonts}
\usepackage{amsmath,epsfig,bm}
\usepackage{graphicx}
\usepackage{color}

\usepackage[setpagesize=false]{hyperref}

\usepackage{tabularx}  

\usepackage{times}

\newcommand{\be}{\begin{equation}}
\newcommand{\ee}{\end{equation}}
\newcommand{\ba}{\begin{eqnarray}}
\newcommand{\ea}{\end{eqnarray}}
\newcommand{\bd}{\begin{displaymath}}
\newcommand{\ed}{\end{displaymath}}

\def\thalf{{\textstyle{\frac{1}{2}}}}
\def\tthalf{{\textstyle{\frac{3}{2}}}}

\def\oneqt{{\textstyle{\frac{1}{4}}}}

\def\Dz{\frac{d}{dz}}
\def\phidot{\dot{\phi}}
\def\phiddot{\ddot{\phi}}
\def\chidot{\dot{\chi}}
\def\chiddot{\ddot{\chi}}
\def\Gdot{\dot{G}}
\def\Gddot{\ddot{G}}
\def\rt6{\sqrt{6}}
\def\mL2{(m_{\phi}L)^2}

\newcommand{\cL}{\mathcal L}
\newcommand{\cS}{\mathcal S}

\begin{document}

\title{{\bf A Dynamical Three-Field AdS/QCD Model}}
\author{Sean P. Bartz and Joseph I. Kapusta}
\affiliation{School of Physics \& Astronomy, University of Minnesota, Minneapolis, MN 55455,USA}

\vspace{.3cm}
\date{\today}

\parindent=20pt

\begin{abstract}
The Anti-de Sitter Space/Conformal Field Theory (AdS/CFT) correspondence may offer new and useful insights into the non-perturbative regime of strongly coupled gauge theories such as Quantum Chromodynamics (QCD). 
We present an AdS/CFT-inspired model that describes the spectra of light mesons. 
The conformal symmetry is broken by a background dilaton field, and chiral symmetry breaking and linear confinement are described by a chiral condensate field. 
These background fields, along with a background glueball condensate field, are derived from a potential. 
We describe the construction of the potential, and the calculation of the meson spectra, which match experimental data well. 
We argue that the presence of the third background field is necessary to properly describe the meson spectra. 
\end{abstract}

\maketitle
\vfill

\section{Introduction}

The Anti-de Sitter Space/Conformal Field Theory (AdS/CFT) correspondence is a useful mathematical tool for the analysis of strongly-coupled gauge theories.
This correspondence establishes a connection between an $d$-dimensional Super-Yang Mills Theory and a weakly-coupled gravitational theory in $d+1$ dimensions \cite{maldacena, Gubser1998, Witten:1998}. 
Calculations that are analytically intractable in the field theory can be related to results from the gravity theory using an effective dictionary developed over the past decade. 
Quantum chromodynamics (QCD) is a strongly-coupled gauge theory at hadronic scales, making it a candidate for the application of the gauge/gravity correspondence.
It is not known whether a gravitational dual to QCD exists, but there has been much work on models that capture its key features.
The bottom-up approach assumes the existence of such a dual, modeling features of QCD by an effective five-dimensional gravity theory.
Linear confinement in QCD sets a scale that is encoded in a cut-off of the fifth dimension in the AdS/QCD model \cite{stephanov-katz-son, DaRold2005}. 
So-called soft-wall models use a dilaton field as an effective cut-off to limit the penetration of the meson fields into the bulk \cite{karch-katz-son-adsqcd}. 
The simplest soft-wall models use a quadratic dilaton to recover the linear Regge trajectories, while models that modify the UV behavior of the dilaton more accurately model the ground state masses \cite{gherghetta-kelley, bartz-pions, Colangelo2008, Cui2013}.

The soft-wall models typically include at least two background fields: the aforementioned dilaton, and a chiral condensate field that corresponds to the chiral symmetry breaking in the gauge theory.
These models use parametrizations for the background dilaton and chiral fields that are not derived as the solution to any equations of motion. 
A well-defined action would provide a set of background equations from which these fields can be derived, and may suggest how the model can be derived from a top-down approach. 
In addition, this action provides access to the thermal properties of the model through perturbation of the geometry \cite{Herzog2007, Bayona2008, Gursoy2008b}.

In this paper, we expand upon previous work to find a suitable potential for the background fields of a soft-wall AdS/QCD model \cite{Batell2008, Springer2010, Gursoy2008a, Gursoy2008b, Csaki2007, Li2013, Li2013a, He2013}. 
After demonstrating the limitations of models including a dilaton and chiral field alone, we suggest the inclusion of a background glueball field. 
We then construct a potential that satisfies the necessary UV and IR limits, and use this potential to generate numerically the background fields and calculate the resulting meson spectra.

\section{Review and Motivation}
\label{secReview}

We assume that four-dimensional QCD can be modeled by the following five-dimensional action, written in the string frame:
\ba
\cS &=&\frac{1}{16\pi G_5} \int d^5x \sqrt{-g} e^{-2\Phi}  \Bigg( R+4\partial_M\Phi\partial^M\Phi \nonumber \\ 
& & \mbox{} - \mathrm{Tr} \left[ |DX|^2  +\partial_M \mathcal{G} \partial^M \mathcal{G} + \frac{1}{2g_5^2} (F_A^2+F_V^2) +V_m(\Phi,X^2, \mathcal{G}) \right]\Bigg) \, .
\label{eqStringAction}
\ea
Here $\Phi$ is the dilaton and the metric is pure AdS, $g_{MN}=z^{-2}\eta_{MN},$ with the AdS curvature defined to be unity.
The constant $g_5^2 = 12\pi^2/N_c$, where $N_c$ is the number of colors.
The covariant derivative is defined as $D_M = \partial_M+i[V_M,X]-i\{A_M,X\}$.
The scalar field $X$, which is dual to the $\bar{q}q$ operator, obtains a $z$-dependent vacuum expectation value (VEV)
\be
\langle X \rangle=\frac{\chi(z)}{2}I \, ,
\ee
where $I$ is the $2 \times 2$ identity matrix.
The glueball field $\mathcal{G}$ similarly obtains a $z$-dependent VEV, $G(z)$.
We examine the background dynamics of the fields
\be
\cS =\frac{1}{16\pi G_5} \int d^5x \sqrt{-g} e^{-2\Phi}  \left(R+4\partial_M\Phi\partial^M\Phi -\thalf\partial_M \chi \partial^M \chi -\thalf\partial_M G \partial^M G -V(\Phi,\chi,G) \right) \, ,
\ee
where $V=\mathrm{Tr}[V_m]$.
The scalar fields $\Phi,\chi,G$ are dimensionless. 

It is easier to search for the background fields in the Einstein frame, where the vacuum action takes the canonical form
\be
\cS_E=\frac{1}{16\pi G_5} \int d^5x \sqrt{-\tilde{g}}\left(\tilde{R}-\thalf\partial_M\phi\partial^M\phi -\thalf\partial_M\chi\partial^M\chi -\thalf\partial_M G \partial^M G - \tilde{V}(\phi,\chi,G)\right) \, .
\label{eq:Einstein}
\ee
The tilde distinguishes the two frames, with $\tilde{V}=e^{4\Phi/3}V,$ and the dilaton is rescaled for a canonical action $\phi=\sqrt{8/3}\Phi$.
The string and Einstein frame metrics are related by the conformal transformation
\be
g_{MN}=e^{2\phi/\sqrt{6}}\tilde{g}_{MN} \, .
\ee

Previous work showed how to construct a potential for a gravity-dilaton-chiral system without the glueball condensate. 
We examine the behavior assuming that the fields have power-law behavior, which is accurate in both the UV and IR limits \cite{Springer2010}. 
One of the equations of motion is independent of the choice of potential,
\be
\chidot^2  = \frac{\rt6}{z^2} \Dz(z^2\phidot) \, . 
\label{twofield}
\ee
To obtain linear confinement, the dilaton should have quadratic behavior in the IR limit, $\phi(z)=\lambda z^2$.
The chiral field should have linear behavior in the IR, $\chi(z)=A z$, where $A$ sets the mass splitting between the axial-vector and vector mesons for large radial quantum numbers $n$. 
This constant mass-splitting at large $n$ occurs because of the non-restoration of chiral symmetry \cite{Shifman-2008}.
Inserting this into (\ref{twofield}), we find that the chiral field behaves as
\be
\chi(z)=6^{3/4}\sqrt{\lambda}z \, ,
\ee
which removes one of the independent parameters of the model in \cite{gherghetta-kelley}. 
Using the phenomenological value of $\lambda$, which determines the slope of the radial Regge trajectories, we find a mass splitting that is much too large.
Because this problem arises in the equation that is independent of the potential, this issue cannot be resolved by the choice of potential in models that do not consider the glueball condensate. 
Models that derive the field behavior using the superpotential method suffer from the same problem.

To resolve this problem, we consider the effects of the glueball condensate $G$ on the background equations. 
This field must be linear in the IR for linear confinement, and behave as $G \sim z^4$ in the UV to match the operator dimension in the AdS/CFT dictionary.

It is noted that the model proposed by Huang and Li \cite{Li2013, Li2013a} accurately represents the non-restoration of chiral symmetry using a model with only two background fields, but their model differs from the work presented here in several respects.
They place the meson fields and chiral dynamics in the open-string sector of the model. 
For linear confinement, this requires that the chiral field approach a constant in the IR, which necessitates a modified metric to obtain the correct chiral dynamics.
Our model allows the metric to remain purely AdS in the string frame.
Finally, they do not determine an explicit form of the potential, which is the central goal of this work.

\section{Construction of Potential}

Consider the action in the Einstein frame (\ref{eq:Einstein}).
To simplify the equations of motion, we use a transformed potential, 
\be
V=e^{-2\phi/\rt6}\tilde{V} \, .
\label{transform}
\ee
This is simply the potential in the string frame.
We re-write it as
\be
V = -12 + 4\sqrt{6}\phi + a_0\phi^2 +\frac{m_X^2}{2}\chi^2 + U \,.
\label{V}
\ee
Here $U$ is more than quadratic in the fields. 
 The AdS/CFT dictionary sets the mass for the fields according to the dimension of the dual operator,
\be
m^2L^2=\Delta(\Delta-4) \,,
\ee
where $L$ is the AdS curvature which we set to unity.  The dimension of the $q\bar{q}$ operator is 3, so $m_X^2 = -3/L^2$.
The dilaton mass is undetermined and is not connected to the dimension of the corresponding operator, as discussed in \cite{Springer2010}.  
It is related to the parameter $a_0$ by $a_0 = \thalf \left[ \mL2-8 \right]$. 
The potential should be an even function of $\chi$. 

The equations of motion can be written as
\be
\chidot^2 + \Gdot^2 = \frac{\rt6}{z^2} \Dz(z^2\phidot) \, ,
\label{C}
\ee
\be
U=\thalf \rt6 z^2 \phiddot - \tthalf (z\phidot)^2 - 3 \rt6 z\phidot 
-4\sqrt{6}\phi - a_0\phi^2 +\tthalf\chi^2 \, ,
\label{U}
\ee
\be
 \frac{\partial U}{\partial \phi}=3z\phidot - 2a_0\phi \, ,
\label{phi}
\ee
\be
 \frac{\partial U}{\partial \chi}
=z^2\chiddot -3z\chidot \left(1+\frac{z\phidot}{\rt6} \right) + 3\chi \, ,
\label{chi}
\ee
\be
 \frac{\partial U}{\partial G}=
z^2\Gddot -3z\Gdot \left(1+\frac{z\phidot}{\rt6} \right) \, .
\label{G}
\ee
We assume that the potential has no explicit dependence on the coordinate $z$,  so the equations \ref{phi}-\ref{G} are not independent, and we can eliminate one. 

\subsection{Infrared Limit}

The requirement of linear confinement requires a solution in the large $z$ limit of the  form
\ba
\phi &=& \lambda z^2 \, , \\
\chi &=& Az \, , \\
G &=& B z \, .
\label{Lz}
\ea
Substitution into (\ref{C}) gives
\be
A^2 + B^2 = 6\rt6 \lambda \, .
\label{Clarge}
\ee
The parameter $\lambda$ is fixed by the slope of the linear trajectory and $A$ is fixed by the axial-vector -- vector mass difference.  
It is useful to write these as
\bd
A = 6^{3/4} \sqrt{\lambda} \cos\theta \, ,
\ed
\be
B = 6^{3/4} \sqrt{\lambda} \sin\theta \, ,
\ee
where $\theta$ now becomes the parameter controlling the axial-vector -- vector mass splitting.
Inserting (\ref{Lz}) into (\ref{U}-\ref{G}) suggests the following terms in our ansatz for the potential
\be
U =  a_1 \phi \chi^2 + a_2 \phi G^2 + a_3 \chi^4 + a_4 G^4 + a_5 \chi^2 G^2 
+ a_6 G^2 \tanh(g\phi) \, .
\ee
We see that there must be a $G^2$ term in the IR limit, but this is forbidden in the weak-field limit because the glueball condensate field is massless. 
To circumvent this, we propose the term $G^2 \tanh(g\phi)$ with $g>0$.  
In the weak field limit this goes to $g\phi G^2$, which is acceptable.  
The $\tanh$ is suggested by \ref{transform}, and it provides a rapid exponential transition from the weak field to the strong field limits that is supported by phenomenology.
By substitution one finds the following constraints on the parameters:
\bd
U \rightarrow 6 + a_0 + 6\rt6 \left( \cos^2 \theta \, a_1 + \sin^2 \theta \, a_2 \right)
\ed
\be
+ 6^3 \left( \cos^4 \theta \, a_3 + \sin^4 \theta \, a_4 + \cos^2 \theta \sin^2 \theta \, a_5 \right) = 0 \, ,
\ee
\be
\frac{\partial U}{\partial \chi} \rightarrow
2a_1 + 24\rt6 \cos^2\theta \, a_3 + 12\rt6 \sin^2\theta \, a_5 + \rt6 = 0 \, ,
\ee
\be
\frac{\partial U}{\partial G} \rightarrow
2a_2 + 24\rt6 \sin^2\theta \, a_4 + 12\rt6 \cos^2\theta \, a_5 + \rt6 = 0 \, ,
\ee
\be
\frac{\partial U}{\partial G} \rightarrow a_6 = - \tthalf \, .
\label{LargeZ2}
\ee
We have chosen to exclude (\ref{phi}) because it is not independent. 
The parameter $a_6$ is determined, and the others will be determined by an examination of the UV limit.

\subsection{Ultraviolet Limit}

Next we look for a solution in the small $z$ limit. 
The AdS/CFT dictionary dictates that the leading-order UV behavior of the chiral and glueball condensate fields is determined by their dimension. 
Note also that we are working in the chiral limit where the quark mass is zero. 
We start by examining only the leading-order terms
\ba
\chi &=& \Sigma_0 z^3 \, ,\\
G &=& G_0 z^4 \, .
\ea
Substitution into (\ref{C}) and imposing the boundary condition $\phi(0)=0$ gives
\be
\phi = \frac{\rt6}{28} \Sigma_0^2 z^6 + \frac{\rt6}{27} G_0^2 z^8 \, .
\label{Sz}
\ee
Using only this leading-order behavior in (\ref{U}-\ref{G}), the system of equations is inconsistent, as there are more equations from matching powers of $z$ than unknown parameters. 

To solve this problem, consider adding a term $\Sigma_n z^n$ to $\chi$.  
Substituting into (\ref{C}) and keeping only the lowest-order cross-term we find the additional term in $\phi$
\be
\Delta \phi = \frac{\rt6 n \Sigma_0 \Sigma_n}{(n+4)(n+3)} z^{n+3} \, .
\ee
From (\ref{U}) we find that
\be
U = -\tthalf (z\phidot)^2 - a_0\phi^2 +3 \frac{n^3 -13n +12}{(n+4)(n+3)} \Sigma_0 \Sigma_n z^{n+3} \, .
\ee
Since the $\phi^2$ terms start out as $z^{12}$, $z^{14}$, $z^{16}$, and so do the terms in the potential, the $n$ can only take the values 9, 11, etc.  
This term contributes only to the equation for $\partial U/\partial \chi$.
\be
\frac{\partial U}{\partial \chi} = -9\Sigma_0 \left( \frac{3}{14} \Sigma_0^2 + \frac{8}{27} G_0^2 z^2 \right) z^9 + (n-3)(n-1) \Sigma_n z^n \, .
\ee
By power counting both $n=9$ and $n=11$ can contribute.  

There could also be higher order terms in $G$ such as $G_m z^m$.  
This leads to the additional term in $\phi$
\be
\Delta \phi = \frac{8 m G_0 G_m}{\rt6 (m+5)(m+4)} z^{m+4} \, .
\ee
It contributes to the equation for $\partial U/\partial G$ as
\be
\frac{\partial U}{\partial G} = -12G_0 \left( \frac{3}{14} \Sigma_0^2 + \frac{8}{27} G_0^2 z^2 \right) z^{10}
+ m (m-4) G_n z^m \, .
\ee
The choice $m=8$ is not possible as there is no term of the same order to balance it.  
Terms with $m=10$ and $m=12$ are possible.  
These new terms cannot affect the equation for $\partial U/\partial \phi$  nor can they contribute to the equation for $\partial U/\partial \chi$.  
Considering higher order terms in both $\chi$ and $G$ leads to
\be
U = -\tthalf (z\phidot)^2 - a_0\phi^2 +3 \frac{n^3 -13n +12}{(n+4)(n+3)} \Sigma_0 \Sigma_n z^{n+3}
+ \frac{4m(m-4)}{m+4} G_0 G_m z^{m+4} \, .
\ee 
The appearance of these terms can be understood by writing the following schematic expansions.
\bd
\chi \sim \Sigma_0 z^3 + \Sigma_0^3 z^9 + G_0^2 \Sigma_0 z^{11} + \cdot\cdot\cdot
\ed
\bd
G \sim G_0 z^4 + \Sigma_0^2 G_0 z^{10} + G_0^3 z^{12} + \cdot\cdot\cdot
\ed
That is, $\chi$ is an odd function of $\Sigma_0$ and $G$ is an odd function of $G_0$.  
These are the symmetries in the equations of motion.  
They also follow the spirit of the AdS/CFT correspondence in terms of the dimensionality of the operators and the powers of $z$.

Including now $m$ = 10 and 12, and $n$ = 9 and 11, we have the following set of equations in the small $z$ limit, where LHS and RHS refer to the left and right sides of the respective equations:
\ba
U_{\rm LHS} &=& 3 \Sigma_0^4 z^{12} \left[ 4 \frac{\Sigma_9}{\Sigma_0^3} - \frac{(54+a_0)}{2^3 \cdot 7^2} \right] \nonumber \\
&+& \frac{1}{7} \Sigma_0^2 G_0^2 z^{14} \left[ 120 \frac{G_{10}}{\Sigma_0^2 G_0} + 120 \frac{\Sigma_{11}}{\Sigma_0 G_0^2} - \frac{(72+a_0)}{9} \right] \nonumber \\
&+& 2G_0^4 z^{16} \left[ 12 \frac{G_{12}}{G_0^3} - \frac{(96+a_0)}{3^5} \right] \, ,
\ea
\ba
U_{\rm RHS} &=& \Sigma_0^4 z^{12} \left[  \frac{\rt6}{28} a_1 + a_3\right] \nonumber \\
&+& \Sigma_0^2 G_0^2 z^{14} \left[ \frac{\rt6}{27} a_1 + \frac{\rt6}{28} (a_2 + g a_6) + a_5 \right] \nonumber \\
&+& G_0^4 z^{16} \left[ \frac{\rt6}{27} (a_2 + g a_6) + a_4 \right] \, .
\ea
\ba
\left(\frac{\partial U}{\partial \chi}\right)_{\rm LHS} &=& 3 \Sigma_0^3 z^9 \left[ -\frac{9}{14} + 16 \frac{\Sigma_9}{\Sigma_0^3} \right]+ 8 \Sigma_0 G_0^2 z^{11} \left[ - \frac{1}{3} + 10 \frac{\Sigma_{11}}{\Sigma_0 G_0^2} \right] \, , \\
\left(\frac{\partial U}{\partial \chi}\right)_{\rm RHS} &=& \Sigma_0^3 z^9 \left[ \frac{\rt6}{14} a_1 + 4 a_3  \right]+ \Sigma_0 G_0^2 z^{11} \left[  \frac{2\rt6}{27} a_1 + 2 a_5 \right] \, .
\ea
\ba
\left(\frac{\partial U}{\partial G}\right)_{\rm LHS} &=& 6 \Sigma_0^2 G_0 z^{10} \left[ -\frac{3}{7} + 10 \frac{G_{10}}{\Sigma_0^2 G_0} \right]
+ 32 G_0^3 z^{12} \left[ - \frac{1}{9} + 3 \frac{G_{12}}{G_0^3} \right] \, , \\
\left(\frac{\partial U}{\partial G}\right)_{\rm RHS} &=& \Sigma_0^2 G_0 z^{10} \left[ \frac{\rt6}{14} (a_2 + g a_6) + 2a_5 \right] \\
&+& G_0^3 z^{12} \left[ \frac{2\rt6}{27}  (a_2 + g a_6) + 4a_4 \right] \, .
\ea

Altogether, from both the UV and IR limits, there are ten independent equations for the twelve parameters $a_0 - a_6$, $\Sigma_9$, $\Sigma_{11}$, $G_{10}$, $G_{12}$, and $g$.  
We take $g$ as the free parameter to use as the rate of transition from small $z$ to large $z$.  
The parameters in the potential are found to be
\ba
a_0 &=&  \frac{3}{2} \frac{1}{6 + \sin^2 \theta}\left[ 120 + 62 \sin^2 \theta + 63 \rt6 g \sin^2 \theta \right] \, , \\
a_1 &=&  -\frac{3\rt6}{4} \frac{1}{6 + \sin^2 \theta}\left[ 12 + 8 \sin^2 \theta + 9 \rt6 g \sin^2 \theta \right] \, , \\
a_2 &=&  -\frac{\rt6}{4} \frac{1}{6 + \sin^2 \theta}\left[ 32 + 24 \sin^2 \theta + 3 \rt6 g(9 \sin^2 \theta - 2) \right] \, ,
\ea
\be
2 a_3 \cos^2 \theta + a_5 \sin^2 \theta = \frac{1}{24} \frac{1}{6 + \sin^2 \theta}\left[ 24 + 22 \sin^2 \theta + 27 \rt6 g \sin^2 \theta \right] \, ,
\ee
\be
2 a_4 \sin^2 \theta + a_5 \cos^2 \theta = \frac{1}{24} \frac{1}{6 + \sin^2 \theta}\left[ 20 +22 \sin^2 \theta   +  3 \rt6 g (9 \sin^2 \theta -2) \right] \, ,
\ee
\be
a_6 = -\frac{3}{2} \, .
\ee
The coefficients $a_0$, $a_1$, $a_2$ and $a_6$ are determined, while there are two equations for the three coefficients $a_3$, $a_4$ and $a_5$.  
That leaves $a_5$ as a free parameter, to be fit numerically, along with $g$, $\theta$, $G_0$, $\Sigma$, and $\lambda$.

\section{Numerical Solution}

Using the potential discussed, we seek a numerical solution that simultaneously satisfies the UV and IR limits. 
We use equations (\ref{C}, \ref{chi}, \ref{G}), which allows for an additional term in the potential, $\Delta U$, such that 
\be
\frac{\partial }{\partial \chi} \Delta U = \frac{\partial }{\partial G}\Delta U = 0 \, ,
\ee
which will be determined from the numerical solution.

The differential equations represent a stiff system, and treatment of the problem as an initial value problem leads to numerical instabilities. 
We treat it instead as a boundary value problem, using Dirichlet boundary conditions at both boundaries. 
A relaxation method is used in combination with input approximations for the background fields, which are then iterated to find a stable solution to the system with the given boundary conditions. 
Because the system is nonlinear, the solution found is not guaranteed to be unique.

The IR boundary is chosen to be sufficiently large to capture the infrared behavior and to give accurate Regge behavior for the large-$n$ radial excitations of the mesons. 
The UV boundary should approach zero, but it cannot reach zero because of the singularity in the equations of motion. 
This becomes a problem because equation (\ref{C}) allows constant and divergent terms 
\be
\Delta \phi(z) = c_1 + c_2 z^{-1} \, .
\ee
Symbolically, these terms can be set to zero by enforcing the Dirichlet boundary condition $\phi(0)=0$, but this is impossible to enforce numerically. 
Creative choice of UV boundary conditions can eliminate one, but not both, of these unwanted terms without affecting the chiral and glueball fields. 
The behavior of the numerical solutions suggests that the desired UV behavior is an unstable solution to the equations, and therefore difficult or impossible to find with this iterative method.

As an alternative to direct solution, we parameterize the fields as follows:
\be
\Psi(z) = \psi(z)_{UV} f(z) + \psi(z)_{IR} \left(1-f(z)\right) \, .
\ee
Here $f(z)$ is some function that transitions smoothly from 1 at small values of $z$ to 0 at large $z$, while $\psi(z)_{xy}$ represents the known UV and IR limits of the fields $\phi, \chi,$ and $G$. 
The switching functions need not be the same for each field. We choose 
\ba
f_\phi(z)&=&e^{-(\beta_1z)^{10}} \, , \\ \label{param1}
f_\chi(z)&=&e^{-(\beta_2z)^4} \, , \\  \label{param2}
f_G(z)&=&e^{-(\beta_3z)^5} \, . \label{param3}
\ea
The powers of the exponential are chosen to be greater than the known power-law behavior of the fields in the UV limit so as to not interfere with this behavior. 
The $\beta_i$ will be determined by numerical fitting.

The chiral condensate $\Sigma$ is set using the Gell-Mann--Oakes--Renner relation:
\be
(m_u+m_d)\Sigma = f_\pi^2 m_\pi^2 \, .
\ee
Using $m_\pi = 139.6$ MeV, $f_\pi = 92 $ MeV, and $m_u+m_d = 7.0 $ MeV yields a value of $\Sigma = (286\, \mathrm{MeV})^3$.

In all, we have eight parameters to be determined numerically. 
The first constraint is to obtain the best global visual fit to the vector and axial-vector meson spectra.   
We do not simply do a chi-squared fitting to the experimental data because the measurement error for the ground state $\rho$ meson is so much smaller than for the others that this would effectively act as the only constraint. 
Second, we seek to minimize the error in the finite-difference approximations to equations $\ref{C}, \ref{chi},$ and $\ref{G}$.  
This is done to an accuracy of one part in $10^4$. 

Three of the parameters are most phenomenologically relevant: $\lambda,$ which controls the slope of the meson spectra in the large-$n$ limit; $\theta,$ which controls the mass splitting between the $a_1$ and $\rho$ mesons at large $n$, and $\beta_2$, which controls the location of the ``bend" in the $a_1$ spectrum.
For each set of these parameters, the other parameters are determined by a routine that minimizes the error in the equations of motion. 
The parameters found are shown in Table \ref{tabParam}.

\begin{table}[htb]
\begin{center}
\begin{tabular}{| l | c || c | c | }
\hline
  $\lambda^{1/2}$ & $304$ MeV & $\beta_1$ & 3.04 GeV  \\
  $G_0^{1/4}$ & 552 MeV & $\beta_2$ & 274 MeV \\
  $ \theta $& 1.44 & $\beta_3$ & 558 MeV \\
  $g $& 3.20 & $a_5$ & 1.63 \\
  \hline
\end{tabular}
\caption{Best fit parameters for the phenomenological model. 
The parameters $\lambda, \theta,$ and $ \beta_2$ are chosen for the best visual fit to the $\rho$ and $a_1$ data, with the rest set by minimizing the error in the equations of motion (\ref{C}), (\ref{chi}-\ref{G}). }
\label{tabParam}
\end{center}
\end{table}

\begin{figure}[ht]
\center{\includegraphics[width=290pt]{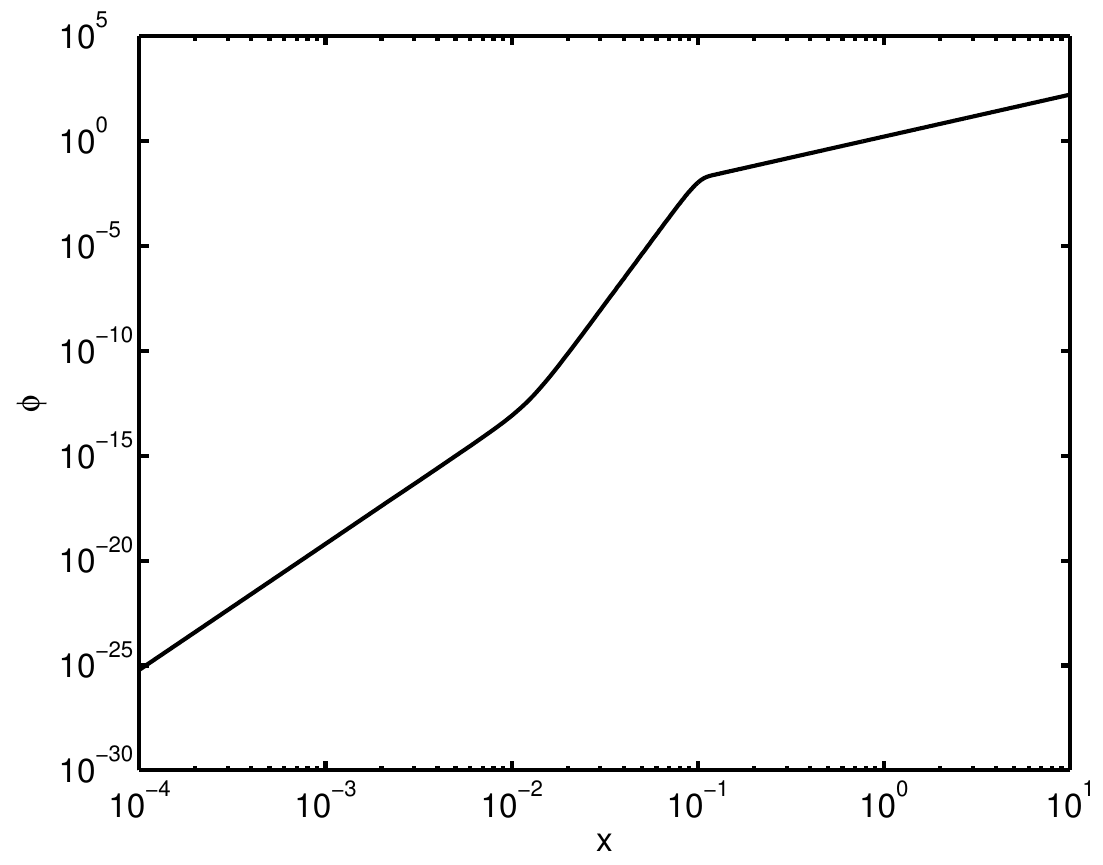}}
\caption{A plot of the dilaton field $\Phi$ generated by the parameterization (\ref{param1}).
The UV and IR asymptotic behavior is apparent.
The coordinate $x$ is a dimensionless re-scaling of the conformal coordinate, $x=\sqrt{\lambda}z$.}
\label{figDilaton}
\end{figure}

\begin{figure}[*hb]
\center{\includegraphics[width=290pt]{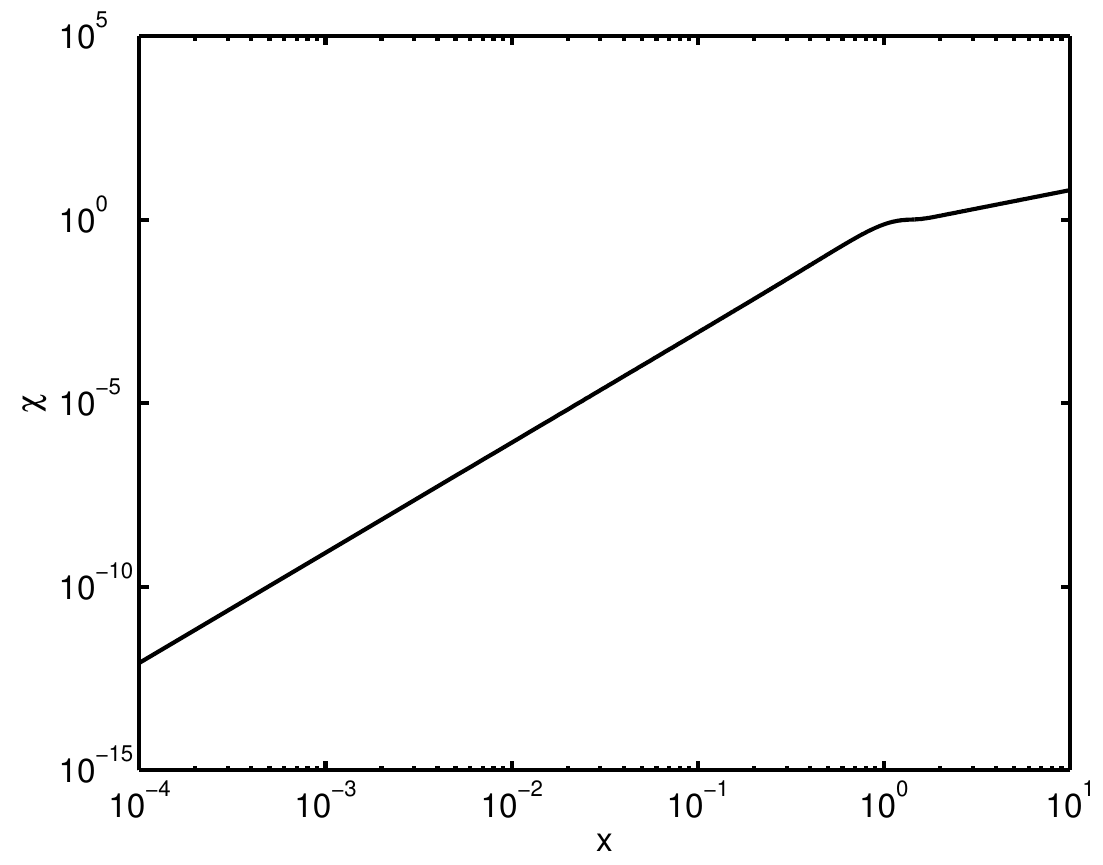}}
\caption{A plot of the chiral field $\chi$ generated by the parameterization (\ref{param2}). 
The UV and IR asymptotic behavior is apparent, with a rapid transition between them.
The coordinate $x$ is a dimensionless re-scaling of the conformal coordinate, $x=\sqrt{\lambda}z$.}
\label{figChiral}
\end{figure}

\begin{figure}[ht]
\center{\includegraphics[width=290pt]{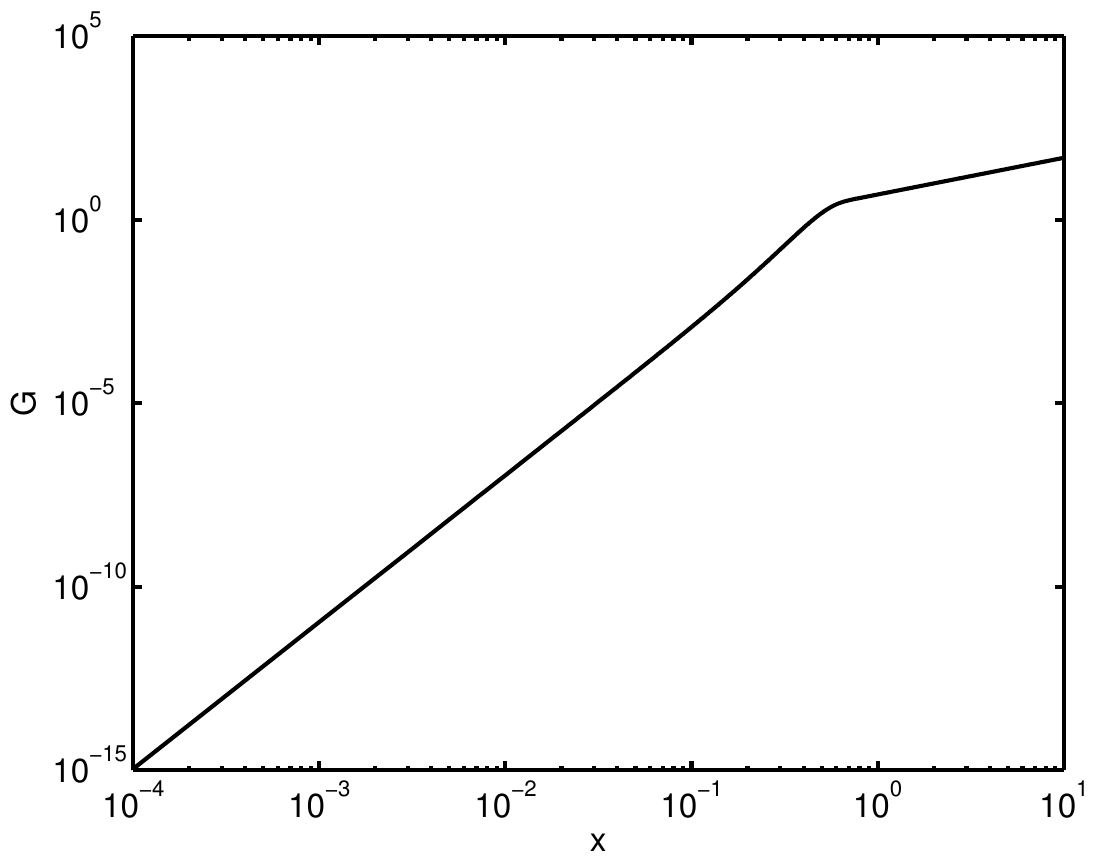}}
\caption{A plot of the glueball field $G$ generated by the parameterization (\ref{param3}).
The UV and IR asymptotic behavior is apparent, with a rapid transition between them.
The coordinate $x$ is a dimensionless re-scaling of the conformal coordinate, $x=\sqrt{\lambda}z$.}
\label{figGlueball}
\end{figure}

The background fields that are obtained from this analysis are shown in Figures \ref{figDilaton}-\ref{figGlueball}. 
The asymptotic power-law behavior of the fields is evident in the linear portions of the log-log scale plots shown.
 The ``transition" behavior is most evident in the dilaton because of the large value of $\beta_1$, which controls the value of $z$ at which the field transitions from the UV limit to the IR limit. 

\clearpage

\nopagebreak

\clearpage

\nopagebreak
\begin{figure}[hb]
\center{\includegraphics[width=285pt]{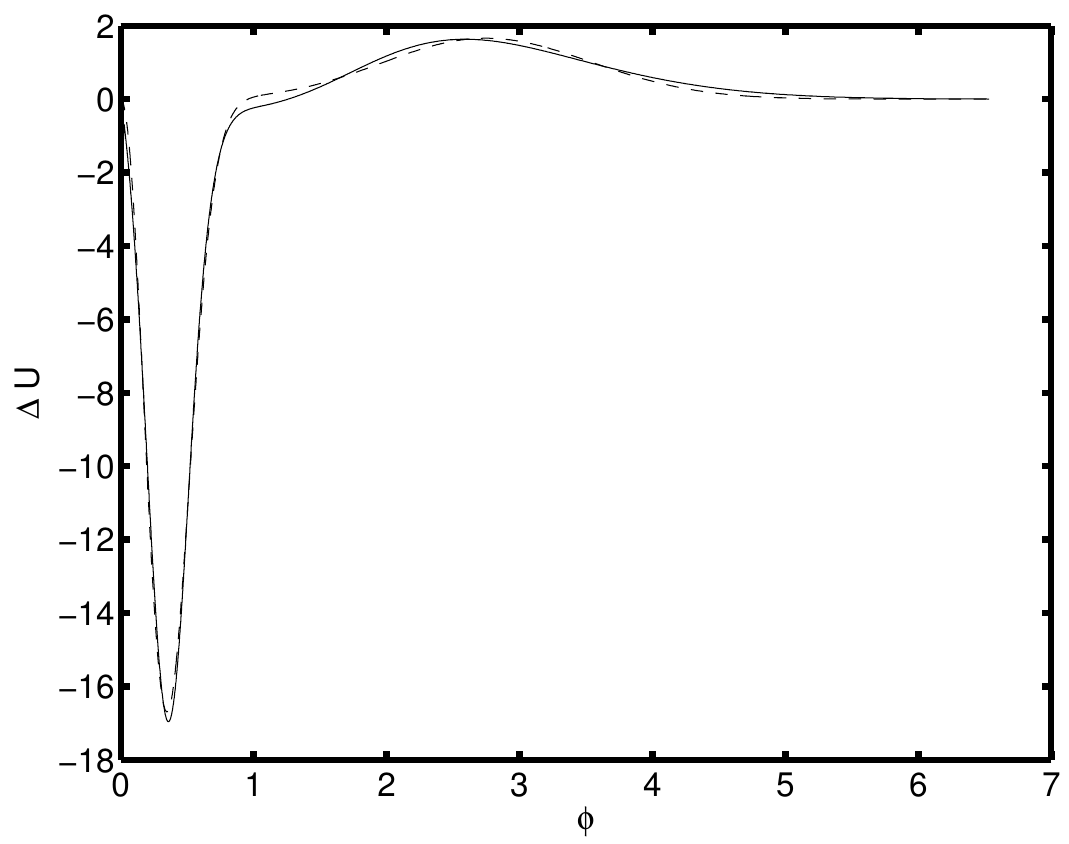}}
\caption{Plot of the ``extra" term in the potential, $\Delta U(\phi)$. The solid line represents the numerical result, while the dashed line is the fitting of (\ref{eqFit}) using the parameters of Table \ref{deltaUfit}.}
\label{figdeltaU}
\end{figure}

\clearpage

We now analyze the ``extra" term in the potential, $\Delta U$. 
We obtain this term numerically by subtracting the right-hand side of \ref{U} from its left-hand side.
This term can be approximated numerically as a function of the dilaton field, 
\be
\Delta U\left(\phi\right) = \alpha_1 \phi^2 e^{-\left(\phi-\gamma_1\right)^2/\delta_1 } +   \alpha_2 \phi^2 e^{-\left(\phi-\gamma_2\right)^2/\delta_2 } \, .
\label{eqFit}
\ee
The best-fit values for these parameters are shown in Table \ref{deltaUfit}.  The $\Delta U$ as a function of $\phi$ is shown in Figure \ref{figdeltaU}.
\begin{table}[htb]
\begin{center}
\begin{tabular}{| l | r || l | r | }
\hline
$\alpha_1$ & $-3.043 \times 10^1$ & $\alpha_2$ & 2.671 $ \times 10^{-4}$ \\
$\gamma_1$ & 7.086 $\times 10^{-5}$ & $\gamma_2$ & 2.213 $ \times 10^{-2}$ \\ 
$\delta_1$ & 9.699 $ \times 10^{-5}$& $\delta_2$ & 1.471 $ \times 10^{-2} $\\ 
  \hline
\end{tabular}
\caption{The dimensionless parameters for the fitting to $\Delta U$.}
\label{deltaUfit}
\end{center}
\end{table}

\section{Vector and Axial-Vector Spectra}

To calculate the spectra of the radial excitations of the mesons, we examine the relevant terms from the string frame action (\ref{eqStringAction}),
\be
\cS_{{\rm meson}}=-\frac{1}{16\pi G_5} \int d^5x \sqrt{-g} e^{-2\Phi}\mathrm{Tr}\left[ \left|DX\right|^2+V_m(\Phi,X^2,\mathcal{G})+\frac{1}{2g_5^2}\left(F_A^2 +F_V^2\right) \right] \, .
\label{eqMesonL}
\ee
The $2 \times 2$ field $X$ contains the scalar and pseudoscalar fields $(S,\pi)$, as well as the VEV.
We will use the exponential representation for the scalar field discussed in \cite{bartz-pions},
\be
X_e = \left( S(x,z)+\frac{\chi(z)}{2}\right)I \, e^{2i\pi^a_e(x,z)t^a},
\ee
where $I$ is the $2\times2$ identity matrix.

\begin{figure}[htb]
\center{\includegraphics[width=300pt]{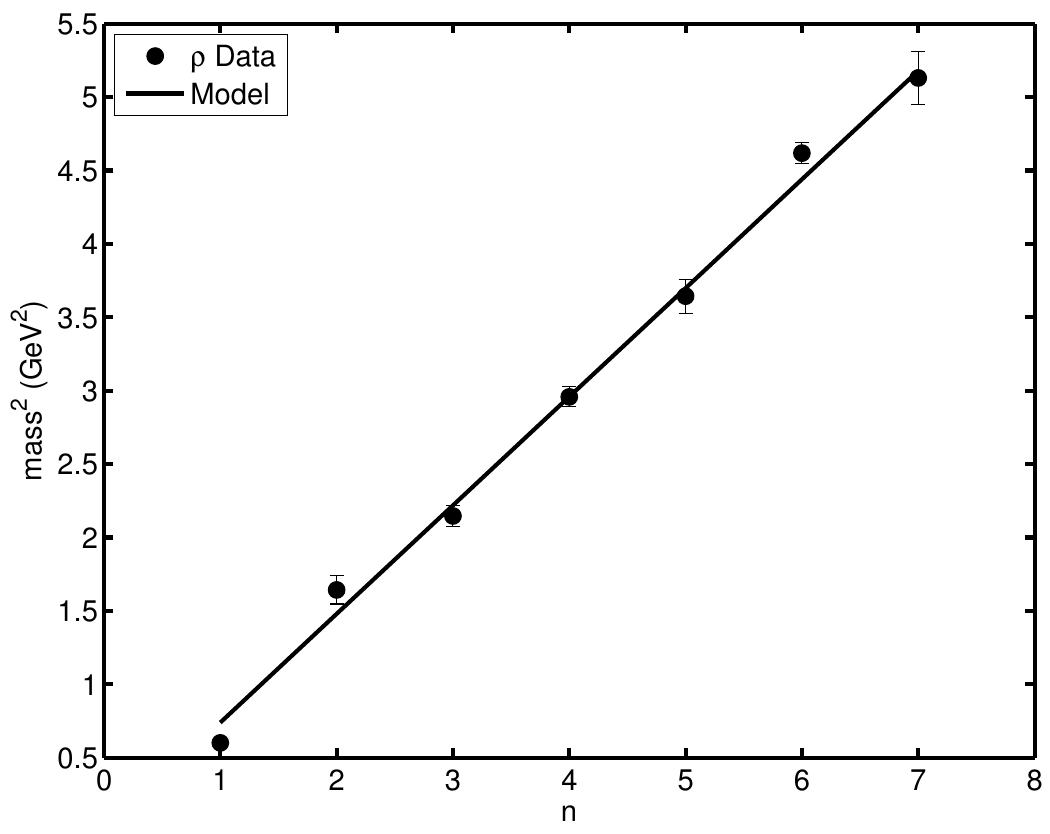}}
\caption{Comparison of the predicted mass eigenvalues for the vector sector with the experimental $\rho$ meson spectrum \cite{PDG}.}
\label{figRho}
\end{figure}

We find the equations of motion for the various meson fields by varying the meson action.
For the vector and axial-vector fields, we  assume that the Kaluza-Klein modes are separable from the 4D parts of the fields.
The equation of motion in the axial gauge $\Psi_5=0$  is given by
\be
-\partial_z^2\Psi_n+\omega'\partial_z\Psi_n +M_\Psi^2(z) \Psi_n=m^2_{\Psi_n}\Psi_n \, ,
\ee
where $\omega=2\Phi(z)+\ln z$. 
The $z$-dependent mass term coefficient $M^2_V=0$  for the vector field, and 
\be
M^2_A=\frac{g_5^2L^2\chi^2}{z^2}
\ee
for the axial-vector field.
The equation can be put in the Schr{\"o}dinger form with the substitution $\Psi_n=e^{\omega/2}\psi_n$, resulting in
\be
-\partial^2_z\psi_n+\left(\oneqt \omega^{'2}-\thalf\omega^{''}+M_\psi^2\right)\psi_n=m^2_{\Psi_n}\psi_n \, .
\ee
These equations are analytically solvable in the IR limit, but full analysis requires the use of a numerical shooting method to find the mass eigenvalues.
This model finds a better phenomenological fit than the results presented in \cite{gherghetta-kelley}, particularly for the ground state $\rho$ meson, as shown in Figure \ref{figRho}. 
The scalar mesons are expected to mix with the scalar glueball field of this model; that analysis is deferred to a future publication. 

\begin{figure}[htb]
\center{\includegraphics[width=300pt]{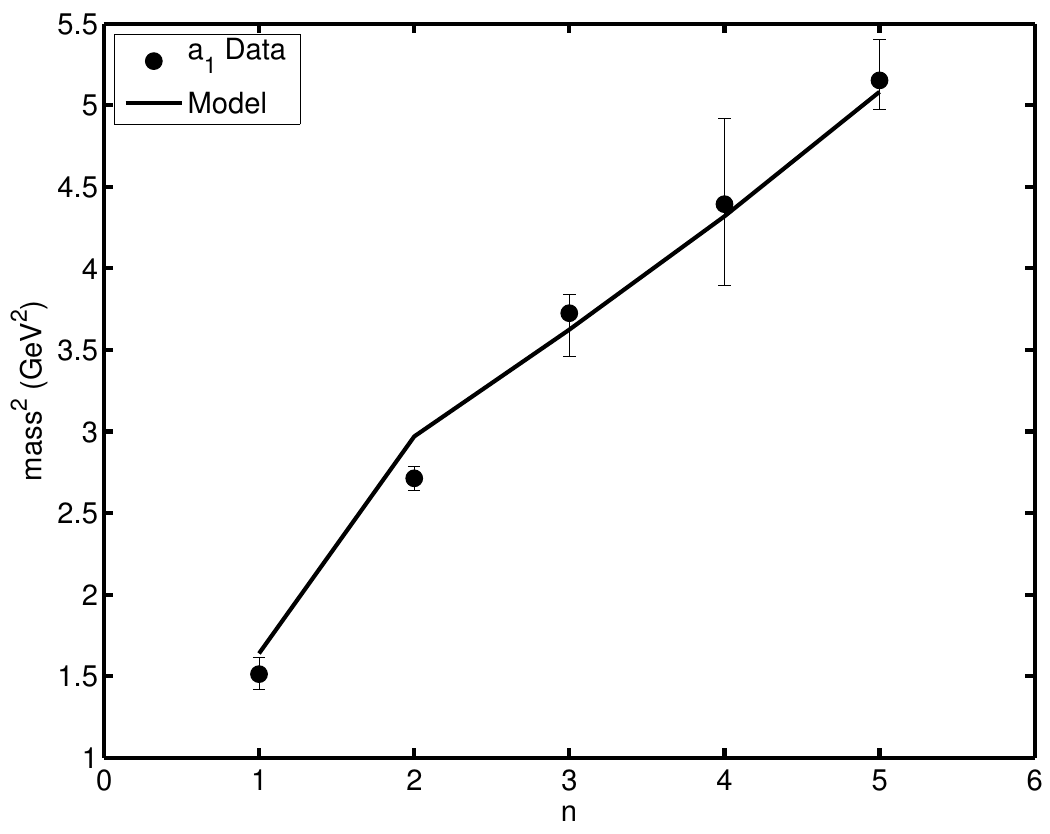}}
\caption{Comparison of the predicted mass eigenvalues for the axial-vector sector with the experimental $a_1$ meson spectrum \cite{PDG}.}
\end{figure}
\nopagebreak

\clearpage

\begin{table}[htb]
\center
\begin{tabular}{| c || c | c  |}
\hline
n & $a_1$ experimental (MeV) & $a_1$ model \\
\hline
1 & 1230$\pm$ 40 &	    	1280	 \\
2 & 1647 $\pm$ 22 & 	1723	 \\
3 & 1930  $^{+30}_{-70}$ & 1904\\
4 & 2096 $\pm$ 122 &      2078	 \\ 
5 & 2270 $^{+55}_{-40}$  & 2254\\
\hline
\end{tabular}
\caption{The experimental \cite{PDG} and predicted values for the masses of the axial-vector mesons.}
\label{tabAxial}
\end{table}

\begin{table}[htb]
\center
\begin{tabular}{| c || c | c  |}
\hline
n & $\rho$ experimental (MeV) & $\rho$ model \\
\hline
1 & 775.5 $\pm$  1 & 860	\\
2 & 1282 $\pm$ 37 & 1216 \\
3 & 1465 $\pm$ 25 & 1489 \\
4 &  1720 $\pm$ 20 & 1720 \\ 
5 &  1909 $\pm$ 30 & 1923 \\
6 &  2149 $\pm$  17& 2107 \\
7 &  2265 $\pm$  40& 2276 \\ 
\hline
\end{tabular}
\caption{The experimental \cite{PDG} and predicted values for the masses of the vector mesons.}
\label{tabRho}
\end{table}

\section{Pseudoscalar Sector}

When using the exponential representation for the scalar field, the terms from the potential do not contribute to the equations of motion for the pion field.
This can be easily seen by noting that $|X_e|^n$ does not contain any terms involving the pion field $\pi_e$ field when $n$ is even. 
We have required the potential to be an even function of $X$, so there are no such terms.
This would seem to suggest that we use the exponential representation to calculate the pion mass spectrum.
However, as noted in \cite{bartz-pions}, $\pi_e$ is extremely sensitive to boundary conditions, and the numerical results are not reliable.
For this reason, we seek to work with an equation of motion written in the linear representation.

For convenience, we begin by deriving the equations of motion in the exponential representation.
Working in the axial gauge $A_z = 0$, we rewrite the axial meson field in terms of its perpendicular and longitudinal components: $A_\mu = A_{\mu\perp} +\partial_\mu \varphi$.
Only the longitudinal component of the axial field, $\varphi$, contributes to the pion equations of motion.
We use (\ref{eqMesonL}), keeping only the relevant terms
\be
\cL = e^{-2\Phi} \sqrt{-g} \left[ \chi^2 (\partial_\mu \pi_e \partial^\mu \pi_e +  \partial_\mu \varphi \partial^\mu \varphi - 2 \partial_\mu \pi \partial^\mu \varphi +  \partial_z \pi_e \partial^z \pi_e)  + \frac{1}{g_5^2}\partial_z \partial_\mu \varphi \partial^z\partial^\mu \varphi \right] \, .
\ee
Varying with respect to $\varphi$ yields
\be
e^{2\Phi} \partial_z \left(\frac{e^{-2\Phi}}{z}\partial_z \varphi \right) + \frac{g_5^2 \chi^2}{z^3}(\pi_e-\varphi)=0 \, ,
\ee
while varying $\pi_e$ gives
\be
\frac{e^{2\Phi} z^3}{\chi^2}\partial_z \left(\frac{e^{-2\Phi}\chi^2}{z^3} \partial_z \pi_e \right) +m_n^2(\pi_e-\varphi) = 0 \, .
\ee

It was shown in \cite{bartz-pions} that the equations of motion are equivalent under the substitution $\pi_e \rightarrow \pi_l/\chi(z)$, so we make the appropriate substitution and expand the equations:
\be
- \varphi'' + \left(2\Phi'+\frac{1}{z}\right)\varphi' = \frac{g_5^2 \chi}{z^2}(\chi \varphi -\pi_l) \, ,
\ee
\be 
-\pi_l'' + \left(2\Phi'+\frac{3}{z}\right)\pi_l' + \left(\chi''-2\chi' \Phi' - \frac{3 \chi'}{z}\right)\frac{\pi_l}{\chi} = m_n^2 (\pi_l - \chi \varphi) \, .
\ee
We can put these equations into Sch{\"o}dinger-like form with the following substitutions:
\ba
\varphi & = & e^{\omega/2}\varphi_n \, , \\
\pi_l & = & e^{\omega_s/2} \pi_n \, , 
\ea
with $\omega =2 \Phi + \ln z$ and $\omega_s = 2 \Phi + 3\ln(z)$.
This yields 
\be
-\varphi_n''+\left(\oneqt \omega'^2-\thalf \omega" +\frac{g_5^2 \chi^2}{z^2}\right)\varphi_n = \frac{g_5^2 \chi}{z} \pi_n \, ,
\ee
\be
-\pi_n''+\left(\oneqt \omega_s'^2-\thalf \omega_s" +\frac{\chi''}{\chi}-\frac{2\chi' \Phi'}{\chi} - \frac{3 \chi'}{z\chi}-m_n^2 \right) \pi_n = -m_n^2 \frac{\chi}{z}\varphi_n \, .
\label{eqPiEOM2}
\ee
The dependence of these equations of motion on the scalar potential can be made explicit by using the background equation for the chiral field, written here in the string frame
\be
z^2\chi'' -3z\chi' \left(1+\frac{z\Phi'}{\rt6} \right) = m_X^2\chi +\frac{\partial U}{\partial \chi} \, .
\ee
Substituting, we can re-write (\ref{eqPiEOM2}) as
\be
-\pi_n''+\left(\oneqt \omega_s'^2-\thalf \omega_s" +\frac{m_X^2}{z^2}+\frac{1}{z^2} \frac{\partial U}{\partial \chi} -m_n^2 \right) \pi_n = -m_n^2 \frac{\chi}{z}\varphi_n \, .
\ee

\begin{figure}[htb]
\center{\includegraphics[width=300pt]{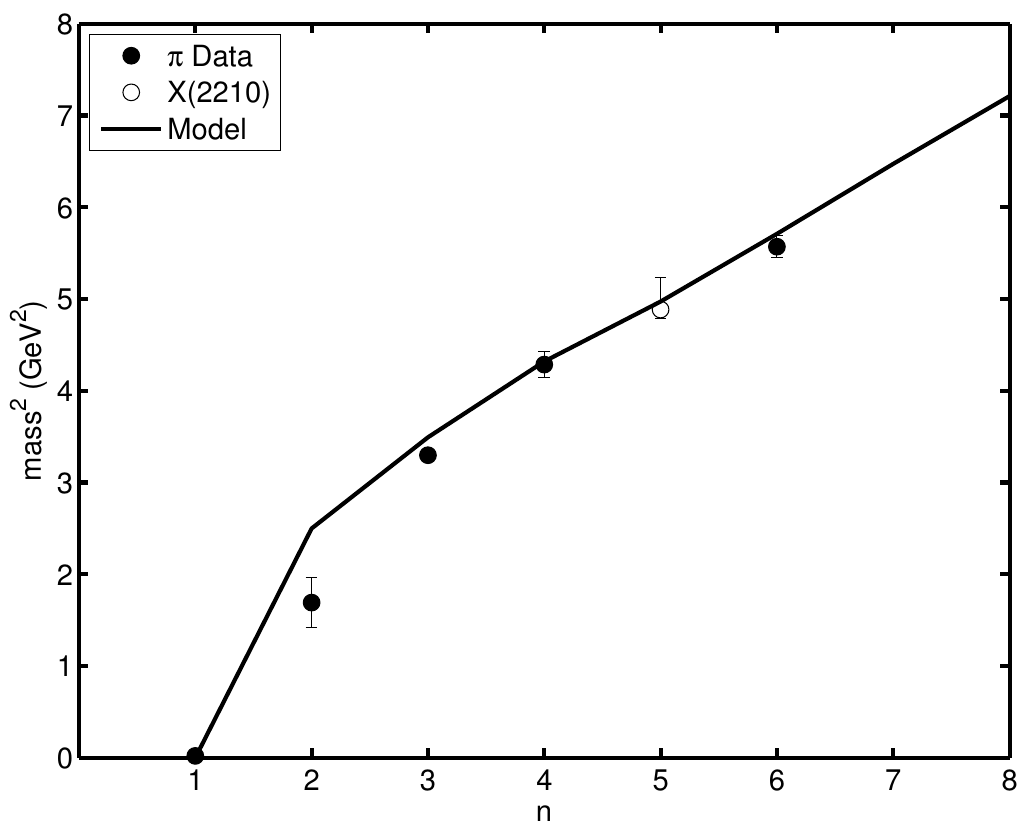}}
\caption{Comparison of the predicted mass eigenvalues for the pseudoscalar sector with the experimental $\pi$ meson spectrum \cite{PDG}.
  The states plotted here with $n=4$ and $n=6$ are identified as radial excitations of the pion only in the further states of the PDG.  
  The unconfirmed state X(2210), with unknown quantum numbers, is plotted here as the $n=5$ state of the pion.}
\label{figPion}
\end{figure}

The results are shown in Figure \ref{figPion} and in Table \ref{tabPion}.  It should be emphasized that all parameters were previously determined, so these are truly predictions of the model.
The states with mass 2070 and 2360 MeV are listed in the PDG as further states, with less certainty assigned to them.
We assume that these should be identified as the $n=4$ and $n=6$ states, leaving a vacancy at $n=5$ for a state still to be observed in future experiments.
On the other hand, the PDG has two further states listed as X(2210) with unknown quantum numbers, either of which could be the $n=5$ state.
We include this state in the figure and in the table, but it should be recognized that nothing in our work depends on this very speculative identification.

\clearpage 

\begin{table}[htb]
\center
\begin{tabular}{| c || c | c  |}
\hline
n & $\pi$ experimental (MeV) & $\pi$ model \\
\hline
1 & 140 &				0 \\
2 & 1300 $\pm$ 100 & 	1580 \\
3 & 1816 $\pm$ 14&		1868 \\
4 & 2070 $\pm$ 35* & 	2078 \\ 
5 & 2210 $^{+79}_{-21} \, \dagger $ &	2230	\\
6 & 2360 $\pm$ 25* & 				2389 \\
7 & -- & 				2544 \\
8 & -- &				2686 \\
\hline
\end{tabular}
\caption{The experimental \cite{PDG} and predicted values for the masses of the pseudoscalar mesons.  
The states marked with an * appear only in the further states of the PDG.  
The state marked with a $\dagger$ is an unconfirmed resonance X(2210) with unknown quantum numbers.  Whether it really represents the $n=5$ state is pure speculation.}
\label{tabPion}
\end{table}

\section{Conclusion}

In this paper we discussed the construction of a potential for the background fields of a soft-wall AdS/QCD model. 
We showed the limitation of a model that contains only the dilaton and chiral condensate fields, and suggested a solution by adding a glueball condensate to the model.
We analytically constructed a general potential $U(\phi,\chi,G)$ that recovers the necessary asymptotic behavior of the background fields.
Using this as a basis, we numerically constructed a potential that solves the selected background equations to within an accuracy of $10^{-4}$. 
There is an additional allowed term in the  potential, $\Delta U(\phi)$, that does not affect the equations that were used in the numerical procedure. 
This term was found numerically, and fit as a function of the dilaton field.
These background fields were then be used to find the spectra of the radial Regge mass spectra of the vector and axial-vector mesons.
The model shows good phenomenological agreement with the experimental data for these spectra.
With the parameters thusly determined, we computed the radial Regge mass spectrum for the pseudoscalar mesons (pions).
Again there was good agreement, except for the most massive state, which perhaps should be identified with the radial quantum number $n=6$ instead of $n=5$.

The potential as constructed here is not guaranteed to be unique.
If a different set of the background equations were chosen, the extra term would be expressed as a function of fields other than the dilaton.
The parameterization in (\ref{param1}-\ref{param3}) could also be chosen differently, resulting in a different potential but making little difference to the resulting meson spectra.
Finally, terms can be added that do not affect the equations of motion at all, namely, terms which satisfy 
\be
\Delta U = \Delta \frac{\partial U}{\partial \phi} = \Delta \frac{\partial U}{\partial \chi} = \Delta \frac{\partial U}{\partial G} =0 \, .
\ee

This work demonstrates the construction of a potential for the background fields of a soft-wall AdS/QCD model that captures several key features of QCD observed through meson spectra.
The radially excited states of the light mesons have linear Regge trajectories.
Chiral symmetry is not restored for highly-excited mesons, as seen in the constant mass-splitting of the vector and axial-vector mesons.
Working as we are in the limit of zero up and down quark masses, the pion is massless.

Future improvements to this model could include incorporating the light quark masses by adding a linear term to the UV limit of the chiral condensate field.
The scalar mesons and glueballs will mix, and this analysis is left for future work.
This potential also opens the possibility of exploring the thermal properties of a model that has the correct chiral symmetry breaking behavior.

\section*{Acknowledgments}
SB thanks Tony Gherghetta and Aleksey Cherman for useful discussions.  
SB also acknowledges Tom Kelley for the development of the code for calculating the pseudoscalar spectrum.
This research is supported by the Department of Energy Office of Science Graduate Fellowship Program (DOE SCGF), made possible in part by the American Recovery and Reinvestment Act of 2009, administered by ORISE-ORAU under contract no. DE-AC05-06OR23100, by the US Department of Energy (DOE) under Grant No. DE-FG02-87ER40328, and by a Doctoral Dissertation Fellowship from the University of Minnesota.


\end{document}